\documentclass[12pt]{article}
\usepackage[margin=1in]{geometry}

\usepackage{hyperref}       
\usepackage{booktabs}       
\usepackage{amsfonts}       
\usepackage{nicefrac}       
\usepackage{microtype}      
\usepackage{graphicx}
\usepackage{moreverb}
\usepackage{amsmath}

\title{A Stochastic Model for Estimating the Number of Offenders and Targets on Snapchat Platform}

\author{
  Vasyl Pihur\\
  Snap, Inc\\
  \texttt{vpihur@snapchat.com} \\
}

\begin{document}
  
\maketitle

\begin{abstract}
Snapchat, like many other social platforms, provides mechanisms for its users to report content and private/public interactions that violate their sense of safety and decency.  From our experience and common sense, we can safely assume that not everybody makes an effort to report, leaving potentially a large number of offending users and content unnoticed. The goal of this work is to directly estimate the probability of someone reporting on Snapchat using current in-app reporting options and, thereby, to provide estimates of the total prevalence (count) of offenders and users subjected to their unwanted, unwelcome or unsafe interactions. 
\end{abstract}



\section*{Introduction}
Snapchat provides in-app tools for its users to report either users or content that are deemed offending, unsafe, annoying or violating Snapchat Terms of Service. Daily, thousands of Snapchat users take advantage of these tools to protect themselves and the platform from unwanted and unwelcome exposures. These are hard, invaluable data, yet it does not provide a view into the full offensive landscape on Snapchat. Without any doubt, not everybody makes the effort to report!

To understand the prevalence of something, one does not need to \emph{identify} every element individually and count them. It is rather impossible to do so in most cases. Often, it is much easier to perform this estimation ``anonymously``, meaning we can estimate the size or cardinality of a set, without knowing what is exactly in it. There are several major techniques and approaches that are based on statistical estimation and sampling theory that allow for this indirect measurement. We will be taking one of these approaches in this work to understand the prevalence of offensive interactions on Snapchat, as well as their reach, in terms of how many users they have affected.

Our in-app reports data provides a \emph{window} into the health of our platform. After all, these are the users who, as of today, actually felt the need to report someone or something. They went through the trouble of interrupting their fun time interacting with their friends and taking a stand. If only all of them have done so. 

We know that there must be a reasonably large number of users who have been subjected to offending content or behaviors, but for one or another reason did not make an effort to report. In this work, we propose a very simple probabilistic model that aims to estimate the proportion of users who did not report (the non-report rate) and provide sensible estimates of their numbers.

Throughout the text, we adopt the following language. Users whose behavior violates other users' sense of safety and decency are called \textbf{offenders} of Snapchat Terms of Service. Users subjected to these violations are called \textbf{targets}. Let $O$ be the total number of offenders and $T$ be the total number of targets on the platform. Let $p$ be the proportion of targets who reported their offenders. All three quantities are unobservable in principle, unless $p = 1$.  

\section*{Notation}
First, we will establish the notation for the following discussion, since it is quite tricky to keep track of all the counts that come into play. 

Let $N$ be a random variable indicating the number of targets that an offender has reached. The probability of reaching exactly $n$ targets is denoted by
$$
P(N = n) = g(n),
$$
where $g(n)$ simply describes the frequencies of offenders with $n \in {1, 2, \ldots}$ targets. 

Note that $g(n)$ is unobservable, since we do not know \emph{a priori} who the offenders are and, even if we did, it would be close to impossible to understand the extent of their reach. But this is exactly the probability mass function that we are after, since it allows us to estimate the number of offenders $O$ and the number of targets $T$. The goal is learn $g(n)$.

Let $K$ be a random variable indicating the number of times each offender is being reported. Let $f(k)$ be the probability mass function summarizing the frequency of how many offenders are reported $k \in \mathbb{N}$ times
$$
P(K = k) = f(k).
$$

It is important to note the difference in the range of $N$ and $K$. $N$ starts at one, since to be an offender, you must have at least one target. $K$, on the other hand, starts at zero, since some offenders never get reported by anyone!

Every day, in our in-app report data, we can observe a random sample from a conditional version of $f(k)$ where $k > 0$. The probability mass function of our actual observed report counts is thus given by
$$
f(k|k>0) = \frac{f(k)}{1 - f(0)}.
$$

Given a random sample from $f(k|k>0)$, our goal is to estimate $g(n)$. $g(n)$ cannot be uniquely identified form $f(k|k>0)$ without making some explicit assumptions on its structure. The problem is essentially ill-defined without further constraints.

\section*{Probabilistic Modeling}
In this section, we will specify our modeling assumptions, justify them through observations on real in-app report data and simulations and present the stochastic generative model for in-app reports data.

\subsection*{Modeling Assumptions}
We make the following two assumptions that allow the estimation of $g(n)$:
\begin{enumerate}
    \item $g(n)$ follows a Power Law distribution with an unknown exponent $s$ and a maximum number of targets $n_{\max}$.
    \item Targets report their offender with an unknown probability $p$, on average, which is modeled using the Binomial distribution.
\end{enumerate}

In the ecology literature, these types of models, called $N$-mixture models \cite{nmixture1, nmixture2, nmixture3}, have been extensively used to provide wild-life population sizes and species abundance estimates. Specific assumption details differ depending on the application, but the overall structure is the same. 

Both assumptions are reasonably easy to justify. The second assumption around the reporting probability $p$ is somewhat obvious. We have a population of targets who are unaware and independent of each other in terms of reporting, and we assume that only $p$ fraction of them reports. $p$ can be interpreted as the proportion of targets who reports their offenders or an average reporting rate across all targets. Each target is likely to have their own sensitivities to different offensive interactions with their own reporting rates. We do not need to worry about these intricacies here, since we only care about the total report counts, which is a sufficient statistic \cite{sufficient} for the overall reporting rate $p$.

To justify the first assumption around the Power Law distribution \cite{powerlaw} on $g(n)$, we will present \emph{two} arguments in this section that will hopefully make it reasonably apparent why such an assumption is warranted. Before we proceed with that, let us formally define the Power Law distribution.

\subsection*{Power Law Distribution}
The Power Law probability mass function is defined on positive integers in the range $[1, n_{\max}]$ as
$$
PL(x, s, n_{\max}) = \frac{x^{-s}}{\sum_{i=1}^{n_{\max}} i^{-s}},
$$
where $n_{\max}$ is the largest number of targets that any offender has.

\begin{figure}
    \centering
    \includegraphics[scale=.6]{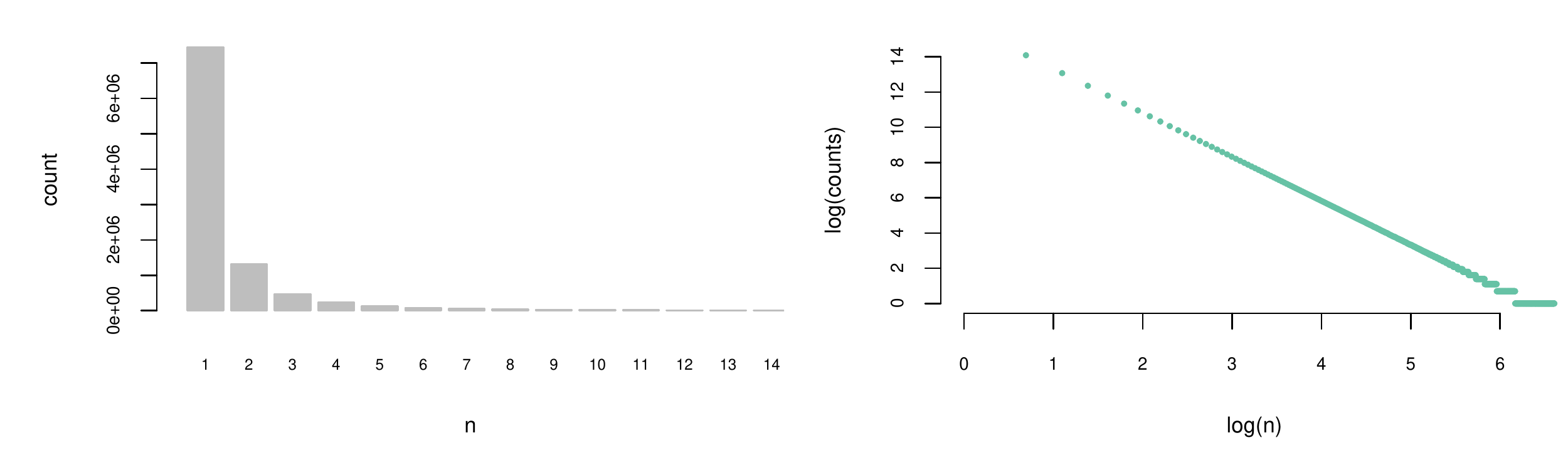}
    \caption{An example of the Power Law distribution with $s = -2.5$ and $n_{\max}=1,000$. The frequency barplot is shown in panel 1 and one can see substantial decay in counts. The log-log plot (both $x$ and $y$-axis are on the logarithmic scale) is shown in panel 2 and clearly follows a straight line, with deviations towards the end due to discrete, integer-valued large $n$'s.}
    \label{fig:example}
\end{figure}

In Figure \ref{fig:example}, we show an example of a Power Law distribution where $s = -2.5$ and $n_{\max} = 1000$. For this set of parameters, by the time one gets to $n=4$, there is very little mass left beyond that point. Perhaps, the most well-known characteristic of the Power Law distribution is the fact that it falls on a straight line on the log-log plot, as shown in the second panel of Figure \ref{fig:example}.

\subsection*{Why Does the Power Law Assumption for $g(n)$ Work?}
Power Law distributions are fascinating and are usually viewed in contrast to the Gaussian (Normal) distributions \cite{normal}. After all, most measurements in daily life do center around their means and fall off symmetrically around it, with most of the mass within three standard deviations. But why is it that some quantities, such as incomes or telephone calls or word frequencies, are not normally distributed?

Beginning in the 1920's when researchers started to notice this phenomenon and were quick to point out that it has to do with the idea of Preferential Attachment\footnote{\href{https://en.wikipedia.org/wiki/Preferential_attachment}{https://en.wikipedia.org/wiki/Preferential\_attachment}}. It commonly exhibits itself as the ``rich getting richer'' process. For any initial distribution of wealth, for example, the steady state that is reached over time is the Power Law distribution. This is due to the fact that everyone, both rich and poor, are assumed to be able to grow their wealth by $X\%$ a year, and not by a fixed dollar amount of $Y$. Whoever had more in the beginning, has an ``unfair'' advantage in \emph{absolute} terms, though the \emph{relative} terms are equal. 

The number of targets per offender, described by $g(n)$ is subject to the same natural forces. The offender who has reached a large number of targets is more likely to continue look for new ones than another random user who has not targeted anyone yet. It is the idea of being ``on a roll''. This phenomenon is closely related to the Chinese restaurant process\footnote{\href{https://en.wikipedia.org/wiki/Chinese_restaurant_process}{https://en.wikipedia.org/wiki/Chinese\_restaurant\_process}}. Imagine the following situation. When a new target is being considered across all Snapchat users, we select a completely new offender with the probability $q$, while the remaining probability $1 - q$ is being split, unequally, among \emph{existing} offenders, and is allocated in proportion to the number of their previous targets. Another way to think about this is the following. Given a list of offender-target pairs, with probability $q$, we select a new offender currently not on the list, match it with a random target and append them to the list. With the remaining probability $1-q$, we pick a pair uniformly at random, extract its offender, match it with a random target and append the new pair back to the list. The more targets the offender have had in the past, the more likely they are to be selected for the new target. The smaller the $q$, the more likely for the same offenders to be targeting more and more users.

\begin{figure}
    \centering
    \includegraphics[scale=.6]{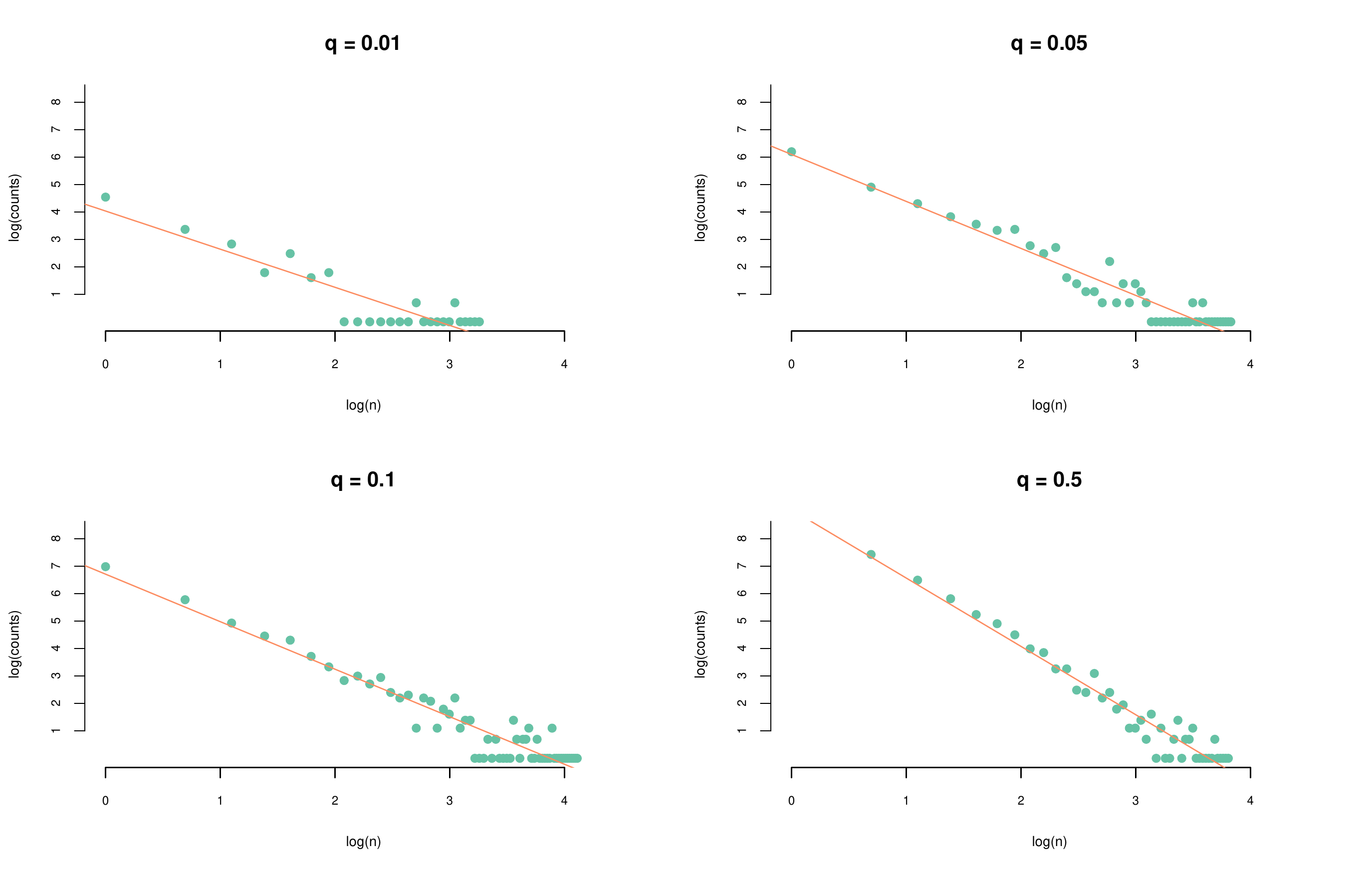}
    \caption{Simulating the offender selection according to the Chinese restaurant process described above. The four panels show simulations for different levels of $q$. In all cases, the log-log plot shows the straight line indicative of the Power Law.}
    \label{fig:chinese}
\end{figure}

We simulate the above process using four different settings for $q$ (the probability of introducing new offenders) and present results in Figure \ref{fig:chinese}. In all cases, regardless of what $q$ is, we obtain the Power Law distribution after running the process 20,000 times. It does not come as a huge surprise that, in every case, we observe a straight line on the log-log plot indicative of the Power Law distribution. 

Offenders who are active are much more likely to continue to be active, while normal users prefer to stay away from offending others most of the time. This dynamic gives rise to the Chinese restaurant process, which in turn, leads directly to the Power Law phenomenon. Therefore, the assumption that the number of targets per offender follows the Power Law distribution is quite reasonable and likely to be expected. Of course, we do not assume a \emph{specific} Power Law distribution for $g(n)$, but let $s$ and $n_{\max}$ to be estimated from the data.

\begin{figure}
    \centering
    \includegraphics[scale=.6]{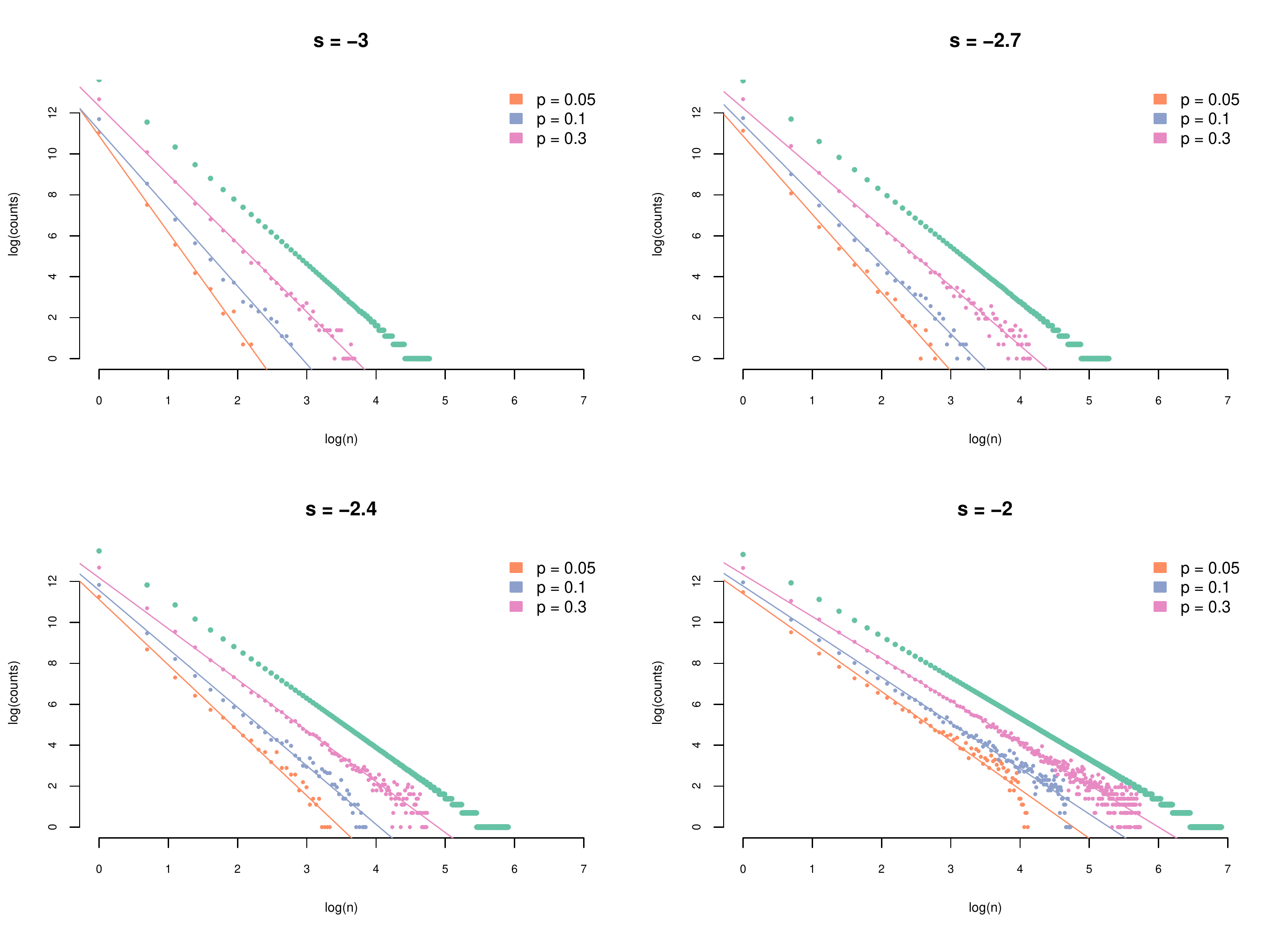}
    \caption{We simulated the generating process under the two assumptions for different values of $s$. We can see that if $g(n)$ follows the Power Law distribution (shown in green), so does $f(k|k>0)$ for different reporting rates $p$, shown in the other three colors with fitted straight lines through them. Since we know that our in-app reports data closely follows a Power Law distribution, we are assured that the Power Law assumption on the unobservable $g(n)$ is a consistent one.}
    \label{fig:power}
\end{figure}

The second argument for the appropriateness of the Power Law distribution comes from the empirical observation of the generative process under the two assumptions of this $N$-mixture model. To get some intuition behind the relationship between $g(n)$ and $f(k|k>0)$, we simulated several scenarios for different values of $s$ and $p$. Results are shown in Figure \ref{fig:power}. We can empirically observe that, if we assume that $g(n)$ follows the Power Law distribution, the Binomial sampling with probability $p$ from that distribution gives rise to something very close to Power Law distribution describing $f(k|k>0)$. It does begin to break down for $s = -2$ as the Power Law for $g(n)$ flattens. It is interesting to note that the counts for $n=1$ are slightly elevated and there appears to be an $s$-shape to the sampled counts, just as we empirically observe in our in-app report data.

Because we empirically observe that our in-app report data follows a Power Law distribution very closely, the assumption that the underlying $g(n)$ is also a Power Law is \emph{consistent} with that observation. It does not mean that $g(n)$ \emph{has} to be Power Law distributed, but that it \emph{can} be without introducing any obvious contradictions to the observed data.

\subsection*{Probabilistic Description of $f(k)$}
$g(n)$ is described by two unknown parameters $s$ and $n_{\max}$. The transition between $g(n)$ and $f(k)$ is governed by the unknown reporting rate $p$ under simple Bernoulli trials, which translates to the Binomial sampling for their sums (counts). Thus, it is possible to specify $f(k)$ in terms of $s$, $n_{\max}$ and $p$. In the literature, this distribution is called the $N$-mixture distribution and is given by
\begin{eqnarray*}
f(k; s, n_{\max}, p) &=& \sum_{n = 1}^{n_{\max}} B(k, n, p) PL(n, s, n_{\max}) \\
           &=& \sum_{n = 1}^{n_{\max}} B(k, n, p) g(n, s, n_{\max}) \\
           &=& \sum_{n = 1}^{n_{\max}} {n \choose k} p^k(1-p)^{n-k} \frac{n^{-s}}{\sum_{i=1}^{n_{\max}} i^{-s}}.
\end{eqnarray*}
where $k \in {0, \ldots, n_{\max}}$.

Here, $B(k, n, p)$ is the probability of getting exactly $k$ reports from $n$ targets using the Binomial distribution. Since, in practice, we do not know how many targets there were, we must compute this for all possible $n$'s and average out the result. This Power Law-Binomial mixture where $n$ varies for each trial is the reason why it is called the $N$-mixture distribution.

If only we knew $s$, $n_{\max}$ and $p$, we could compute the expected distribution of in-app reports $f(k)$. This is where probability theory ends and statistical inference begins. Because we observe a truncated version of $f(k)$, we can reverse the process and estimate those three parameters, which, in turn, gives us $g(n)$.

\section*{Estimation of $s$, $n_{\max}$, and $p$}
Let $k_1, k_2, \ldots, k_M$ be the observed in-app report counts indicating the number of offenders reported once, two times and so forth. The largest observed number of reports per offender is denoted by $M$. We will call these the \emph{observed} report counts. 

Using the Maximum Likelihood estimation (MLE) method \cite{casella}, we will maximize the log-likelihood function to obtain a set of parameter estimates $\hat{s}$, $\hat{n}_{max}$, and $\hat{p}$ that make the observed data, $k_1, k_2, \ldots, k_M$, the most likely among all sets of such parameter estimates.

The log-likelihood function is tightly linked to the $f(k; s, n_{\max}, p)$ $N$-mixture density described above and is given by
$$
ll(s, n_{\max}, p; k_1, k_2, \ldots, k_M) = \sum_{i=1}^{M} k_i \log[{f(i; s, n_{\max}, p)}].
$$

We use numerical optimization routines to maximize this function with respect to $s$, $n_{\max}$, and $p$ and obtain their MLE estimates. To understand how reliable our parameter estimates are, we use a parametric Bootstrap technique \cite{bootstrap} to estimate their variance. 

\begin{figure}
    \centering
    \includegraphics[scale=.6]{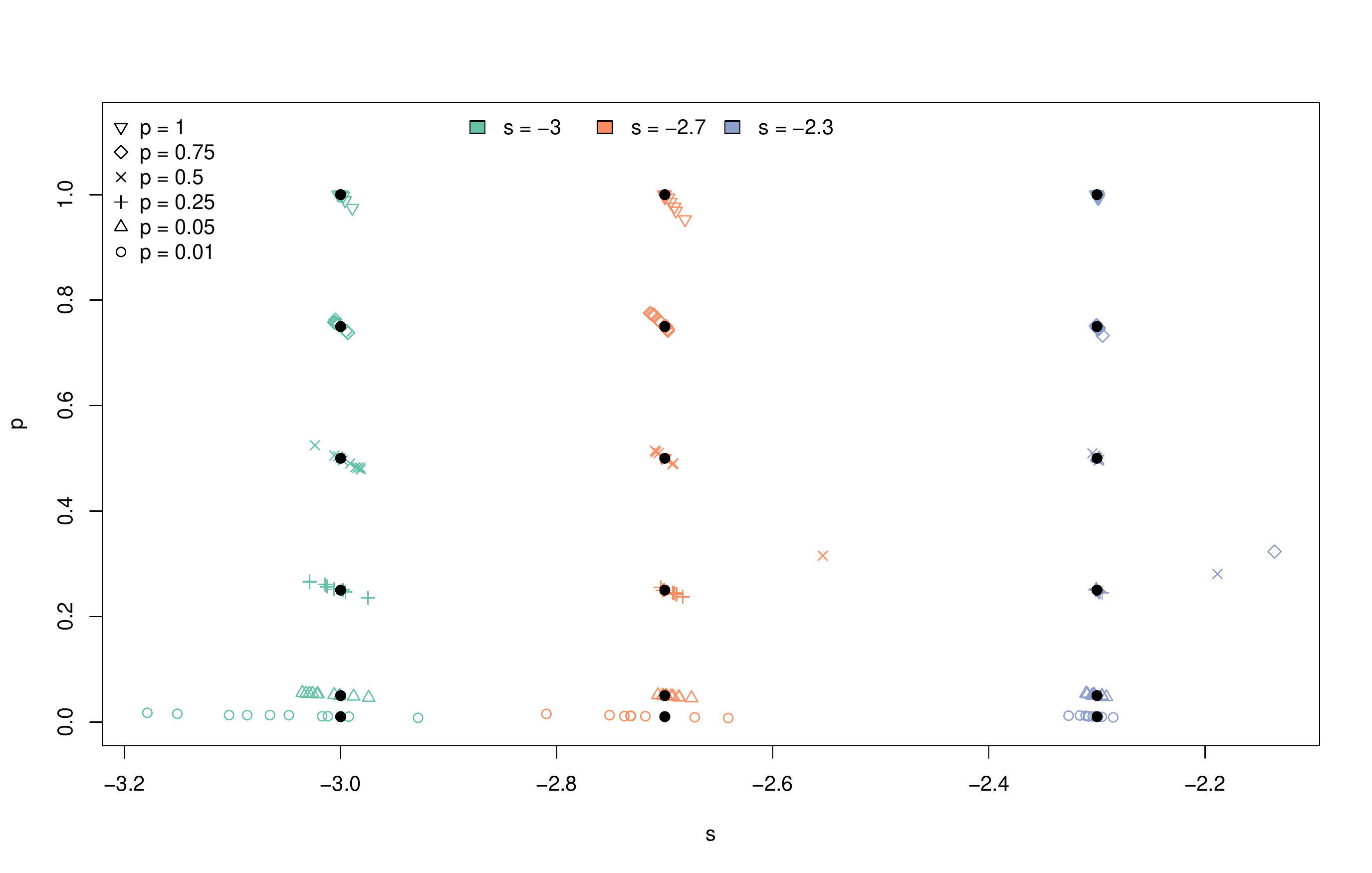}
    \caption{Estimation of $s$ and $p$ in 10 different simulation runs. The true values are denoted by the black dots, while the estimates are shown in different colors and shapes, depending on the combination of true $s$ and $p$ parameters. Estimation is quite accurate for $p > .01$. There can be observed some uncertainty in estimating parameters towards the bottom left corner, where the Power Law distribution is steep and $p$ is small, meaning we get few reports and therefore not much data to work with.}
    \label{fig:sp}
\end{figure}

\subsection*{Estimation of $O$ and $T$}
To estimate $O$, the total number of offenders, we first estimate the proportion of offenders that were never reported, given the estimate of $s$, $n_{\max}$ and $p$
$$
P(K = 0) = f(0; \hat{s}, \hat{n}_{max}, \hat{p}) = \sum_{n = 1}^{\hat{n}_{max}} B(0; n, \hat{p}) PL(n; \hat{s}, \hat{n}_{max}) = \sum_{n = 1}^{\hat{n}_{max}} B(0; n, \hat{p}) g(n; \hat{s}, \hat{n}_{max}).
$$

Since $(1 - P(K = 0))O$ is equal to the number of reported offenders $R$, we get
$$
\hat{O} = \frac{R}{1 - P(K = 0)}.
$$

The estimate of $T$ can now be simply obtained by
$$
\hat{T} = \hat{O} \sum_{n = 1}^{\hat{n}_{max}} ng(n, \hat{s}, \hat{n}_{max}),
$$
where we take the number of offenders with one target multiplied by 1 and add them to the number of offenders with two targets multiplied by 2 and so on.

\begin{figure}[!h]
    \centering
    \includegraphics[scale=.6]{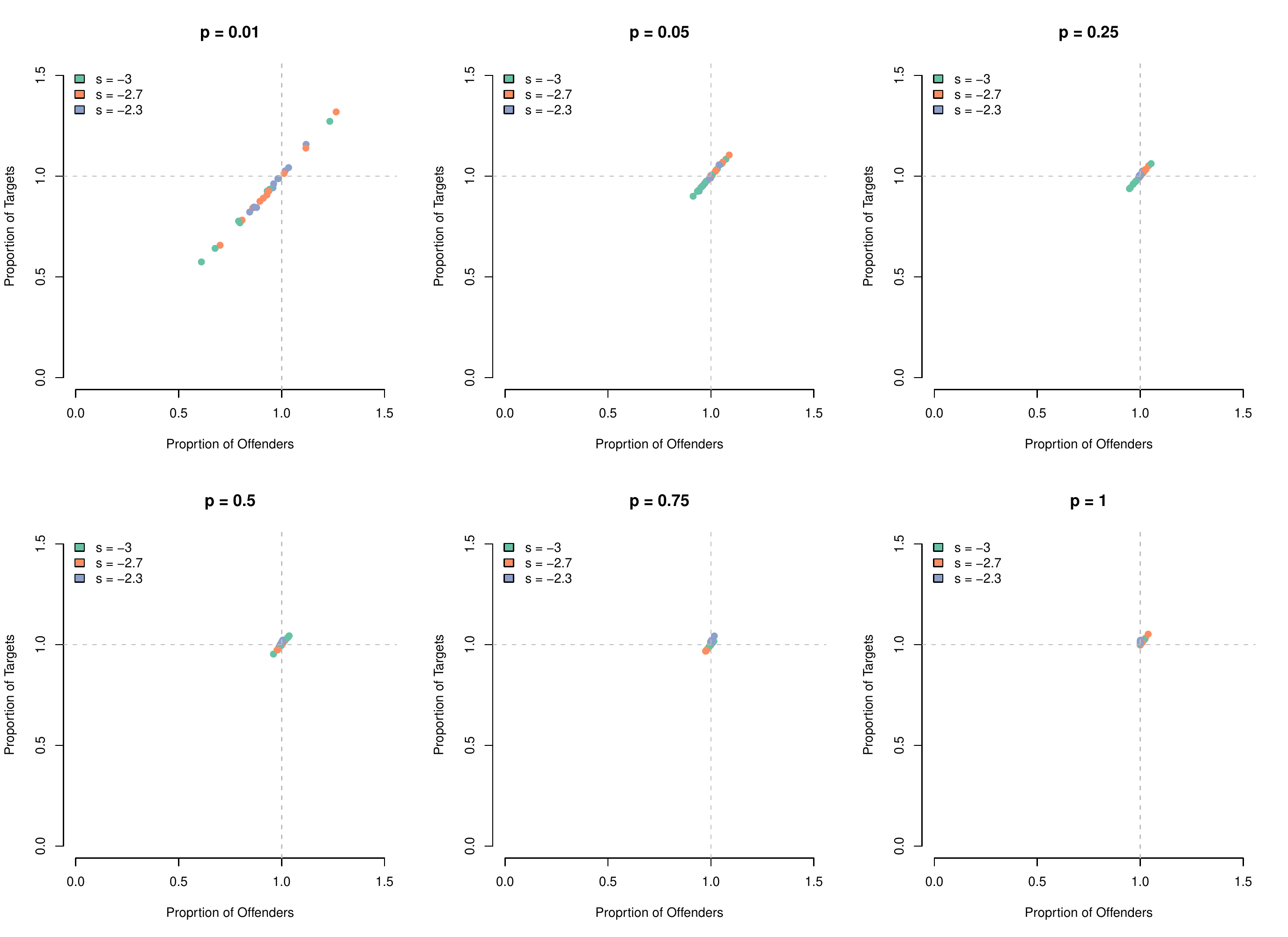}
    \caption{Estimation of $O$ and $T$ in the same 10 simulation runs as in Figure \ref{fig:sp}. Estimates here are normalized by their true value and are equal to them when their ratio's are 1 (on both $x$ and $y$-axes). We again see increased variation for $p=0.01$, but otherwise, estimates are centered around their true values with a reasonable amount of uncertainty.}
    \label{fig:ot}
\end{figure}

\section*{Simulation results}
To verify that our estimation procedure works as intended, we ran 10 simulations for different combinations of $s$ and $p$ and a fixed $n_{\max} = 20,000$. Estimates for $s$ and $p$ over all 10 simulation replica are shown in Figure \ref{fig:sp}. For $p = 0.01$, the estimates have some noticeable variance, while for other combinations, they are quite close to the true values denoted by black dots.

Since we are not directly interested in $s$ and $p$ for their own sake, we would like to estimate $O$ and $T$, the number of offenders and their targets. Simulations results for both $O$ and $T$ in terms of their relative error (divided by true values of $O$ and $T$) are shown in Figure \ref{fig:ot}. Ideally, we would like to have estimates falling as close as possible to 1 on both axes, indicating their equality to the true values. 

For $p = 0.01$, again, we see quite a lot of variation, due to the variation in estimates of $s$ and $p$ that we saw in Figure \ref{fig:sp}. For other values of $p$, depending on $s$, the estimates are close to their true values and cluster around them.

\section*{Discussion}
Modeling prevalence is not easy. After all, we are after quantities which, in principle, cannot be directly observed. Statisticians have spent decades thinking about these problems and developed numerous solutions to tackle them \cite{capture, efron}. $N$-mixture models are commonly used for prevalence estimation and clearly provide reasonable estimates for estimating the population of offenders and targets on Snapchat platform. 

One of my all time favorite statistical papers is one by Efron \cite{efron}, in which he estimates the number of words Shakespeare knew, but did not use in his works. As you can see, estimating unreported offenders is not that much different.

\bibliographystyle{unsrt}  
\bibliography{references} 

\begin{thebibliography}{10}

\bibitem{nmixture1}
Gentile~Francesco Ficetola, Benedetta Barzaghi, Andrea Melotto, Martina Muraro,
  Enrico Lunghi, Claudia Canedoli, Elia Lo~Parrino, Veronica Nanni, Iolanda
  Silva-Rocha, Arianna Urso, Miguel Carretero, Daniele Salvi, Stefano Scali,
  Giorgio Scarì, Roberta Pennati, Franco Andreone, and Raoul Manenti.
\newblock $n$-mixture models reliably estimate the abundance of small
  vertebrates.
\newblock {\em Scientific Reports}, 8, 12 2018.

\bibitem{nmixture2}
J.~Andrew Royle.
\newblock $n$-mixture models for estimating population size from spatially
  replicated counts.
\newblock {\em Biometrics}, 60:108--15, 04 2004.

\bibitem{nmixture3}
Liana Joseph, Che Elkin, Tara Martin, and Hugh Possingham.
\newblock Modeling abundance using $n$-mixture models: The importance of
  considering ecological mechanisms.
\newblock {\em Ecological Applications}, 19, 05 2009.

\bibitem{sufficient}
E.L. Lehmann and G.~Casella.
\newblock {\em {Theory of Point Estimation}}.
\newblock Springer Verlag, 1998.

\bibitem{powerlaw}
MEJ Newman.
\newblock Power laws, pareto distributions and zipf's law.
\newblock {\em Contemporary Physics}, 46(5):323--351, 2005.

\bibitem{normal}
Aidan Lyon.
\newblock {Why are Normal Distributions Normal?}
\newblock {\em The British Journal for the Philosophy of Science},
  65(3):621--649, 09 2013.

\bibitem{casella}
G.~Casella and R.~L. Berger.
\newblock {\em Statistical Inference}.
\newblock Wadsworth and Brooks/Cole, Pacific Grove, CA, 1990.

\bibitem{bootstrap}
B.~Efron and R.~Tibshirani.
\newblock Bootstrap methods for standard errors, confidence intervals, and
  other measures of statistical accuracy.
\newblock {\em Statist. Sci.}, 1(1):54--75, 02 1986.

\bibitem{capture}
Anne Chao, P.~Tsay, Sheng-Hsiang Lin, Wen-Yi Shau, and Day-Yu Chao.
\newblock The applications of capture-recapture models to epidemiological data.
\newblock {\em Statistics in Medicine}, 20:3123 -- 3157, 10 2001.

\bibitem{efron}
B.~Efron and R.~Thisted.
\newblock {Estimating the number of unseen species: How many words did
  Shakespeare know?}
\newblock {\em Biometrika}, 63(3):435--447, 1976.

\end{thebibliography}



\end{document}